\documentclass[final,5p,times,twocolumn]{elsarticle}

\usepackage{subfigure}
\usepackage{lineno,hyperref}

\journal{NIM-A}

\begin{document}

\begin{frontmatter}

\title{DAMIC-M Experiment: Thick, Silicon CCDs to search for Light Dark Matter}

\author[ifca]{N. Castell\'o-Mor\corref{thecorrespondingauthor}}\ead{castello@ifca.unican.es}
\address[ifca]{Instituto de Fisica de Cantabria (Universidad de Cantabria/CSIC), Av. de los Castros s/n 39005 Santander, Spain}
\cortext[thecorrespondingauthor]{Corresponding author. Tel.: +41 22 76 71657}
\author{on behalf of the DAMIC-M Collaboration}
%

\begin{abstract}
    This report presents an overview of the unconventional use of charge-coupled devices
    (CCDs) to search for Dark Matter (DM). The DArk Matter in CCDs (DAMIC
    Experiment) employs the bulk silicon of thick, fully-depleted CCDs as a target
    for ionization signals produced by interations of particle dark matter from the galactic 
    halo. The DAMIC collaboration has engaged in an extensive campaign of characterization 
    efforts to understand the response of these CCDs to low-energy nuclear recoils and their
    unique capabilities, including the use of high spatial resolution for both the
    rejection and study of backgrounds. The preliminary results of DAMIC prove the
    performance of the detector, provide measurements of the background contamination and
    demonstrate the potentiality for DM searches, with only $\sim$40 grams of
    detector mass. 
    The next phase of the experiment, DAMIC-M (DArk Matter in CCDs at Modane), will consist of a
    kg-sized detector, implementing the most massive CCDs ever built. These CCDs will feature
    sub-electron noise and will be deployed in a low-radioactivity environment at the \emph{Laboratoire
    Souterrain de Modane} in France.
\end{abstract}

\begin{keyword}
dark matter detectors\sep
solid state detectors\sep
very-low energy charge particle detectors\sep
CCD
\end{keyword}

\end{frontmatter}


\section{State-of-the-art in DM searches}
\label{sec:Overwiew}

The mistery of dark matter (DM) is one of the most fundamental questions in physics.
Significant efforts have been made to understand the nature of dark matter and theories have been
formulated to explain its existence. Some of these include \emph{Modified Newtonian Dynamics} in the
context of Einstein's General Relativity~\cite{dm_mnd}, existence of yet undetected fundamental
particles with all sort of possible signatures in the context of Particle Physics~\cite{dm_ph} and
Dark Fluid theoris~\cite{dm_fm}.
A well-motivated theoretical model is that DM is composed of a new class of particle(s) 
that was
also produced in an early phase of our universe and interacts (in some ``unknown'' way)
with ordinary matter to dramatically influence the shape of the universe as it is. For
instance, Weakly Interacting Massive Particles (WIMPs), the main focus of the vast
majority of the DM detectors, were produced together with Standard Model (SM) particles in
the hot bath of the early universe, ultimately escaping thermal equilibrium. 
No technique (scintillation crystals~\cite{dm_sc1,dm_sc2,dm_sc3}, noble liquids~\cite{dm_nl1,
dm_nl2, dm_nl3}, bubble chambers~\cite{dm_nc}, cryogenic calorimeters~\cite{dm_cc1,dm_cc2}) has been
successful yet in the effort to detect the low-energy nuclear recoils induced by the interactions of
these theorized particles. The nature of the DM, so far elusive, constitutes one of the most fundamental 
open questions in Physics.

Hidden Sector (HS) particles have
been proposed by the international community as an alternative approach to go beyond the 
WIMP paradigm~\cite{dm_hs}. In
this scenario, DM is made of particles from one of the many ``hidden sectors'' that are
thought to exist outside of the ``visible sector'' (made of ordinary matter) that
encompasses our entire visible world.
Whilst the nuclear recoil induced by light DM is albeit undetectable, energy
transfer in the scattering with electrons or the absorption of a dark photon are much more
efficient~\cite{dm_ese}, allowing DM direct detection experiments to probe as low as ~eV.

Independently of the theoretical motivations, it is important to recognize that current
experiments have limited sensitivity to DM-electron interactions, and a light DM particle
may have well escaped detection. Most of the interactions result in the production of few
charges, requiring the \emph{detector to be able to resolve individual electrons}. An
ubiquitous challenge for DM experiments is also different sources of background (natural 
radioactivity,
airbone radon, neutrons $\alpha$ particles, neutrinos, etc.) which must be really low
for a signal to be recognizable. 
The sensitivity of Si based detectors are limited by the dark
current. A low dark current is a prerequisite to building a detector to search for light DM using
DM-electron interactions.


The best limits on the DM mass are reached with the noble liquid technique with XENON10
data~\cite{xenon10}, but the results are limited by background-induced noise and
were not improved with the tenfold increase in the mass of XENON100. SuperCDMS will
perhaps reach single-electron sensitivity by operating in high-voltage
mode\cite{superCDMS}, however, the
required improvements in phonon resolution and leakage current have not been demonstrated
(so far). 

In this context, innovative technology of a single-electron detection, already demonstrated in CCDs,
will enable DAMIC-M to achieve unprecedented sensitivity to the DM hidden sector.
DAMIC-M capitalizes on the DAMIC experience at SNOLAB (see sec. \ref{sec:damic}) and, at the 
same time, greatly improves in
sensitivity by further innovating the detector technology (see section \ref{sec:damicm}).
In fact, the measurement and mitigation of $^{32}$Si and tritium that will be achieved with 
DAMIC-M are a necessary step to demonstrate the feasibility of a next-generation detector 
aiming to reach the neutrino floor.

\section{CCDs as a Dark Matter detector}
\label{sec:damic}

Recent advances in CCD technology due to the increase in the purity of the silicon have
allowed the fabrication of thicker devices, e.g. 250~$\mu$m-thick used by the Dark Energy
Survey\cite{dm_DES} for efficient detection of near-infrared light from astrophysical
objects. Motivated by its potential, Lawrence Berkeley National Laboratory
developed a high-resistivity silicon CCD for the DAMIC experiment
with the idea to increase sensitivity to search for DM particles. This new prototype of
CCD achieved a record thickness of 675~$\mu$m with an area of 6cm$\times$6cm and a mass
of 5.8 grams. Furthermore, the CCD also presents high spatial resolution and an excellent
energy response in very effective background identification techniques. All this makes
the DAMIC CCDs a well-suited detector to identify and suppress radioactive background.
DAMIC is an initiative to search for dark matter through direct search based on these
technology. It employs the bulk silicon of the  
CCD as a target for interactions of particle DM~\cite{dm_damic}. Ionization signals
may be produced in the active bulk of the device by recoiling nuclei or electrons
following the scattering of a dark matter particle. By virtue of the low readout noise of
the CCD technology, and the relatively low mass of the silicon nucleus, DAMIC is particularly 
sensitive to the signal from low mass WIMPs (2--10 GeV/c$^{2}$), which induce nuclear 
recoils of keV-scale energies (bottom panel Fig.~\ref{fig:DMExpSens}, violete-dotted line).

DAMIC CCDs were fabricated from n-type, high-resistivity silicon
wafers, and are fully depleted (i.e., active over their full volume) by applying a potential 
($\geq 40$V) to a thin back-side
contact. Each CCD is epoxied onto a silicon backing, together with a flex cable that is
wire bonded to the
CCD and provides the voltage biases, clocks and video signals required for its
operation. These components are supported by a copper frame to complete the CCD module.
The modules are installed into slots of a copper box that is cooled to $\sim$130K inside a
vacuum chamber.

The response of DAMIC CCD to ionization radiation has been extensively characterized in
the laboratory. The linear response of the amplifier has been demonstrated with optical
photons for ionization signals as small as 10~$e^{-}$, and with
mono-energetic X-ray and gamm ray sources for energies in the range $0.5-60$keV$_{ee}$
(where 3.8 eV$_{ee}=1e^{-}$) \cite{dm_calibration}.
The energy scale for recoiling nuclei, which produce a smaller electron-hole pairs than a recoiling
electron of the same kinetic energy, was calibrated with neutron sources~\cite{al_cal}.
DAMIC CCDs present some unique properties when is compared to other dark matter detectors:

\emph{1. An unprecedented charge resolution} of a pixel charge r.m.s. 
noise of $\sim2~$e$^{-}$, dominated by the noise of the readout
amplifier. This allows for the positive identification of as little as 40~eV of
ionization energy deposited in a pixel. DAMIC-M will have a much smaller noise
allowing for high-resolution detection of a single electron, being sensitive to
extremely small energy transfers from a DM-electron interaction.

\begin{figure}[!ht]
    \centering
    \includegraphics[width=\linewidth]{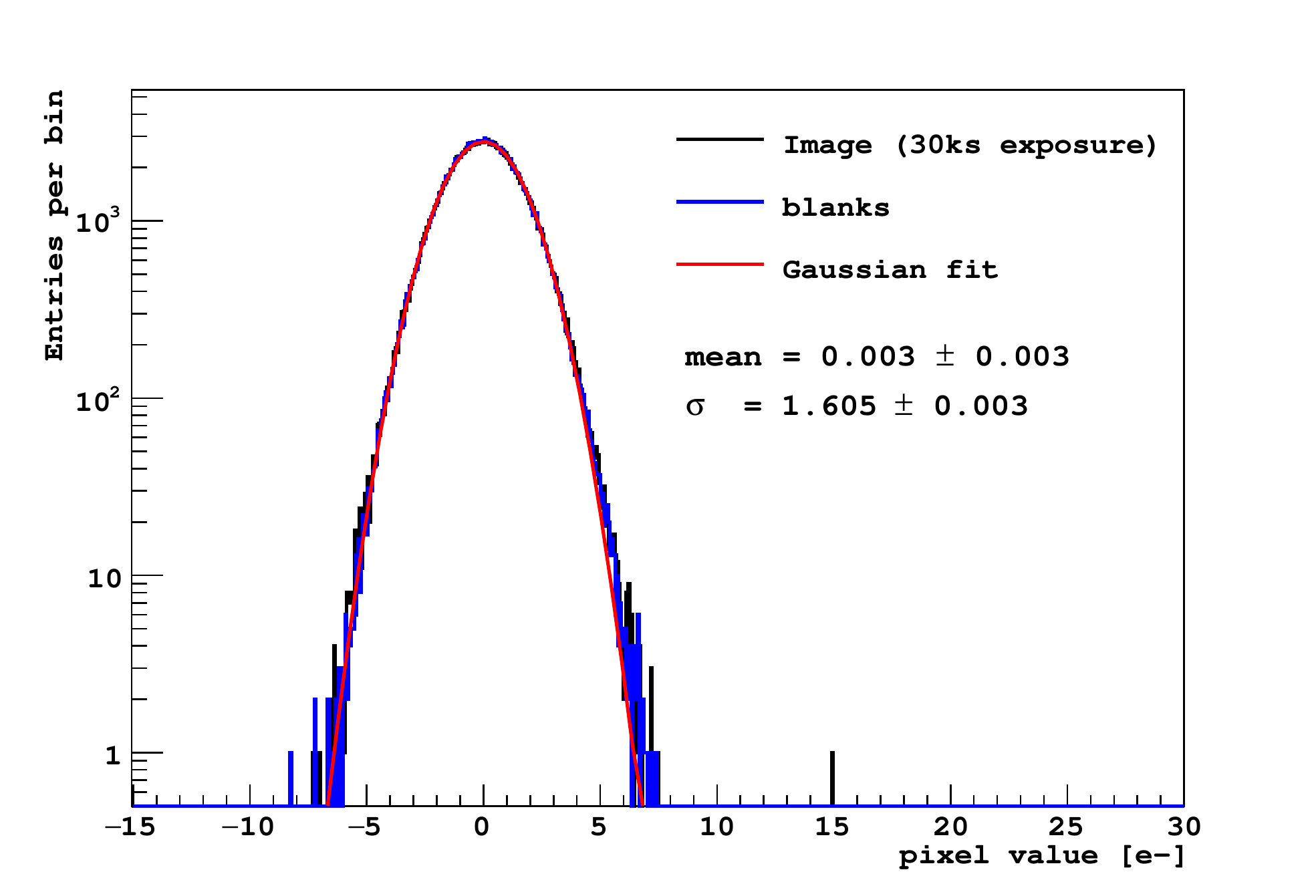}
    \caption{Distribution of pixel values in blanks images (white noise) and in 8 hours exposure
        (white noise, leakage current and sigma) when operating at 140 K.
    }
    \label{fig:lowestDC}
\end{figure}

\emph{2. The lowest leakage current ever measured in a silicon detector} of a level of
4$e^{-}$/mm$^2$/day. This extremely low dark current allowed to place experimental 
constraints on dark matter
interactions that produce as little as a single electron, improving the ionization 
threshold over previous dark matter searches by an order of magnitude.
Figure~\ref{fig:lowestDC} shows the observed pixel distribution, consistent with
the expected distribution of leakage current convolved with a white pixel readout
noise of 1.6$e^{-}$. This capability allows DAMIC CCDs technology to place the most stringent
direct-detection on hidden-photon dark matter in the galactic halo with masses
$3-12$~eV$c^{-2}$\cite{dm_thp}.

\begin{figure}[htb]
    \centering
        \includegraphics[width=\linewidth]{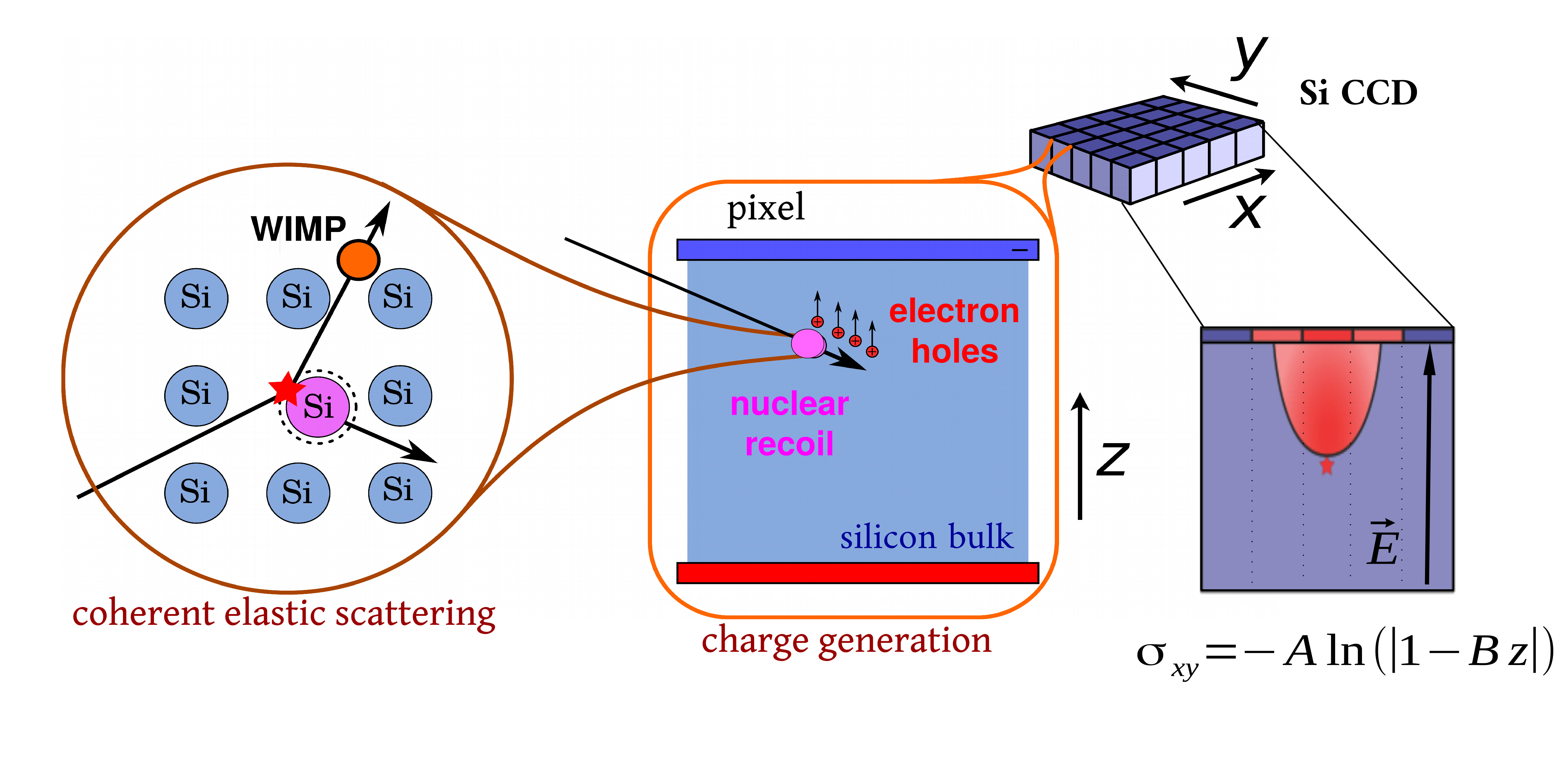}\\
        \includegraphics[width=0.6\linewidth,height=0.25\textwidth]{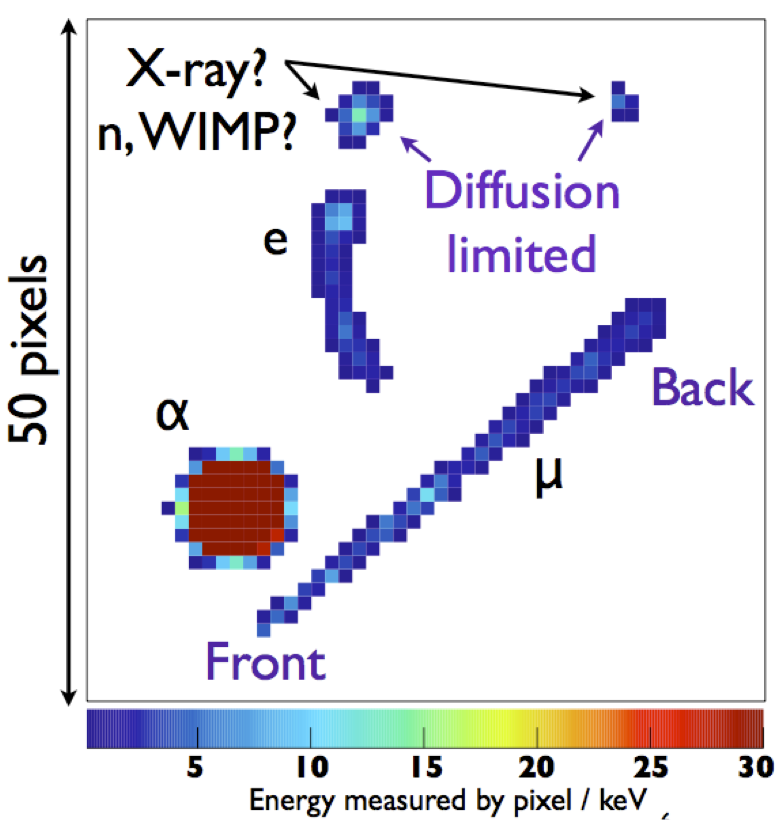}
        \caption{\emph{Top left to right:} Representation of the WIMP-nucleon scattering and of the charge diffusion by
        a point-like ionization event in the CCD bulk. The $x-y$ coordinates give the
        position in the CCD whereas the lateral spread positively correlates to the depth
        of the energy deposit. The diffusion model has been tested with data from
        radioactive sources and cosmic muons. \emph{Bottom:} Signatures of different
        ionizing particles in a CCD: a straight track (cosmic ray muon), large blob (alpha
        particle), ``worm'' (straggling electron) and small round clusters (low-energy X-ray,
        nuclear recoil, DM candidate).}
   \label{fig:ccd}
\end{figure}

\emph{3. An exquisite spatial resolution and 3D reconstruction}. The principle of DM detection 
with a CCD is illustrated in Figure~\ref{fig:ccd}. Ionization charge produced in the 
substrate, through absorption or a
nuclear/electronic recoil, is drifted towards the pixel gate along the direction of the
electric field ($z$ axis) and collected on the pixel array ($x-y$ plane), where it is held
in place until the readout. Because of thermal motion, the ionized charge diffuses transversely 
with respect to the electric field direction as it is drifted, with a spatial variance
($\sigma_{xy}$) that is proportional to the transit time (i.e. the depth, $z$, of the
interaction point). Hence, there is a positive correlation between the lateral 
diffusion of the
collected charge on the pixel array and the depth of the interaction, which allows for the
reconstruction in three dimension of the location of energy depositions in the
bulk of the device, as well as the identification of particle types based on the cluster pattern
(bottom panel on Figure~\ref{fig:ccd}).

\begin{figure*}[htbp]
    \centering
    \includegraphics[width=0.95\linewidth]{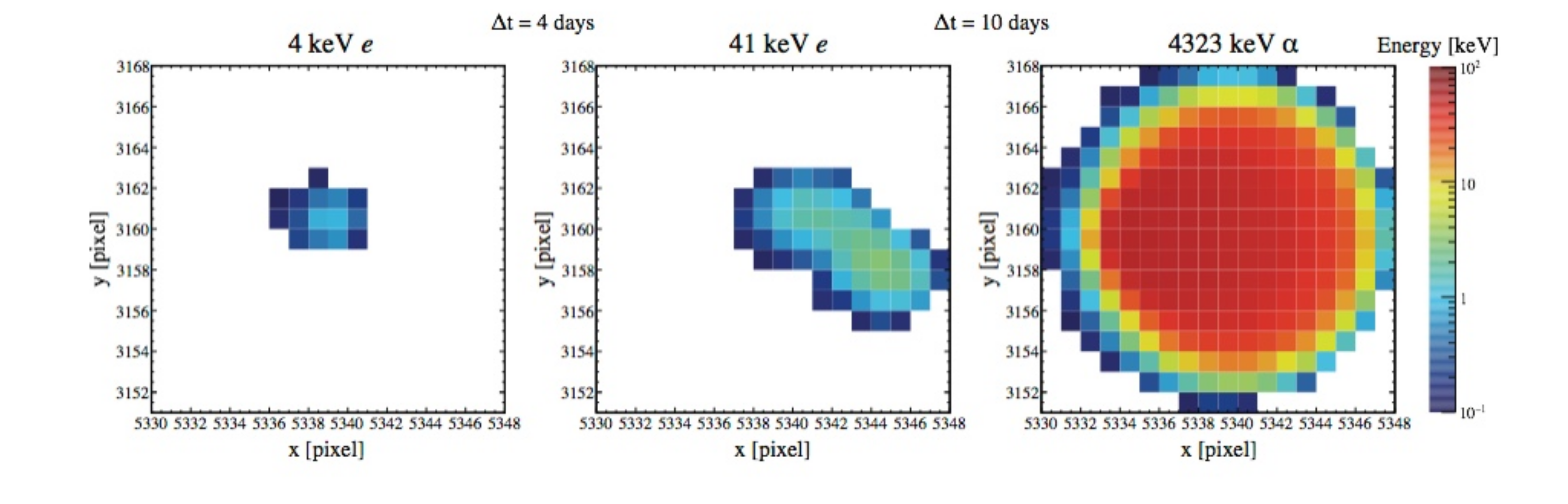}
    \caption{An example of the truly unique capability of DAMIC on background rejections. The three
        cluster (two electrons and one alpha) detected in the same location but separated in time by
        several days. The observation is consistent with the decay chain of a single $^{210}$Pb nucleus 
        on the CCD surface. A product of the radon decay chain, $^{210}$Pb is a major source of
    surface background in DM experiments caused by material exposure to air.}
    \label{fig:decaychain}
\end{figure*}

\emph{4. Background identification and rejection}. A truly powerful capability of DAMIC is that
background can be identified and rejected as spatially correlated events occurring
at different times. An example is shown in Figure~\ref{fig:decaychain}, with three
clusters (two electrons and one alpha) detected in the same location but separated
in time by several days. The probability for this to occur by change is
negligible. The exquisite spatial resolution and the 3D reconstruction are the basis for the 
rejection of background events produced by low energy gammas and electrons on the surface 
of the CCD, as well as, for the characterization of the radioactive background on the surface and 
in the bulk of the CCD. 

\section{Prospects of DAMIC-M}
\label{sec:damicm}

DAMIC-M, which is the successor to the DAMIC experiment has received international recognition and
has been supported by the European Research Council (ERC-Advanced grant). The new detector will be a low-background 
50-CCD array with a mass of 1~kg and a detection threshold of sub-electron. 
several improvements in the detector design, construction materials and CCD
packaging have been foreseen in order to progress further in the search for low-energy dark matter
particles, including the GeV-scale WIMPs, the hidden-photon, and to prove a large region of
parameter space for dark matter particles in the hidden sector. The most relevant features of the
new DAMIC-M CCDs are listed below.

\begin{itemize}
    \item It will contain the most massive ever built CCD (20 grams), three times more
        massive than those at SNOLAB. R\&D on device packaging and handling already started in order
        to avoid mechanical stresses on the CCDs during cooldown as this may cause unwanted induced
        charges to appear an possibly even damage the device.

    \item An important improvement for DAMIC-M will be the implementation for ``skipper'' 
        amplifiers on large area 1-mm-thick 36~Mpixel CCDs. 
        The skipper amplifiers perform a large number of uncorrelated
        measurements of
        the charge collected by each CCD pixel, significantly decreasing the pixel noise by
        averaging over a large number of samples. The single electron response of the
        skipper amplifier has already been demonstrated with a 200 $\mu$m-thick 3.6~Mpixel
        CCD~\cite{dm_skipper}, where a readout noise of 0.07~$e^{-}$ was achieved with 4000
            samples per pixel.
        The single-electron resolution greatly will simplify 
        the calibration since the number of collected electrons per pixel will be counted, 
        providing a direct measurement of the calibration constant.

   \item DAMIC CCD readout will differ from those at SNOLAB, where CCDs are read out every
       8 hours. In DAMIC, the resolution on the pixel charge is dominated by the readout
       noise, and long exposures are preferable to limit the number of pixels being
       read out in the experiment's live time. The readout noise is negligible in the
       skipper CCD, and the leakage current of the device becomes the limiting factor.
       Thus, continuous readout through four skipper amplifiers will be used to minimize 
       the accumulation of charge from
       leakage current in a pixel before it is read out. 

   \item A decrease of the radioactive background to a level of $\sim0.1 \rm{keV}^{-1}
        \rm{kg}^{-1}\rm{day}^{-1}$ will also be necessary. This will require
        improvements in the design of the detector array and in the handling and packaging
        of the devices to mitigate surface backgrounds from $^{210}$Pb. Also, careful
        selection of construction materials and procedures and minimizing the exposure of the
        components to cosmic rays will be implemented to minimize the activation of
        $^{3}$H in the silicon target, which is expected to be the dominant background.
\end{itemize}

\begin{figure}[ht]
    \centering
    \includegraphics[width=0.6\linewidth]{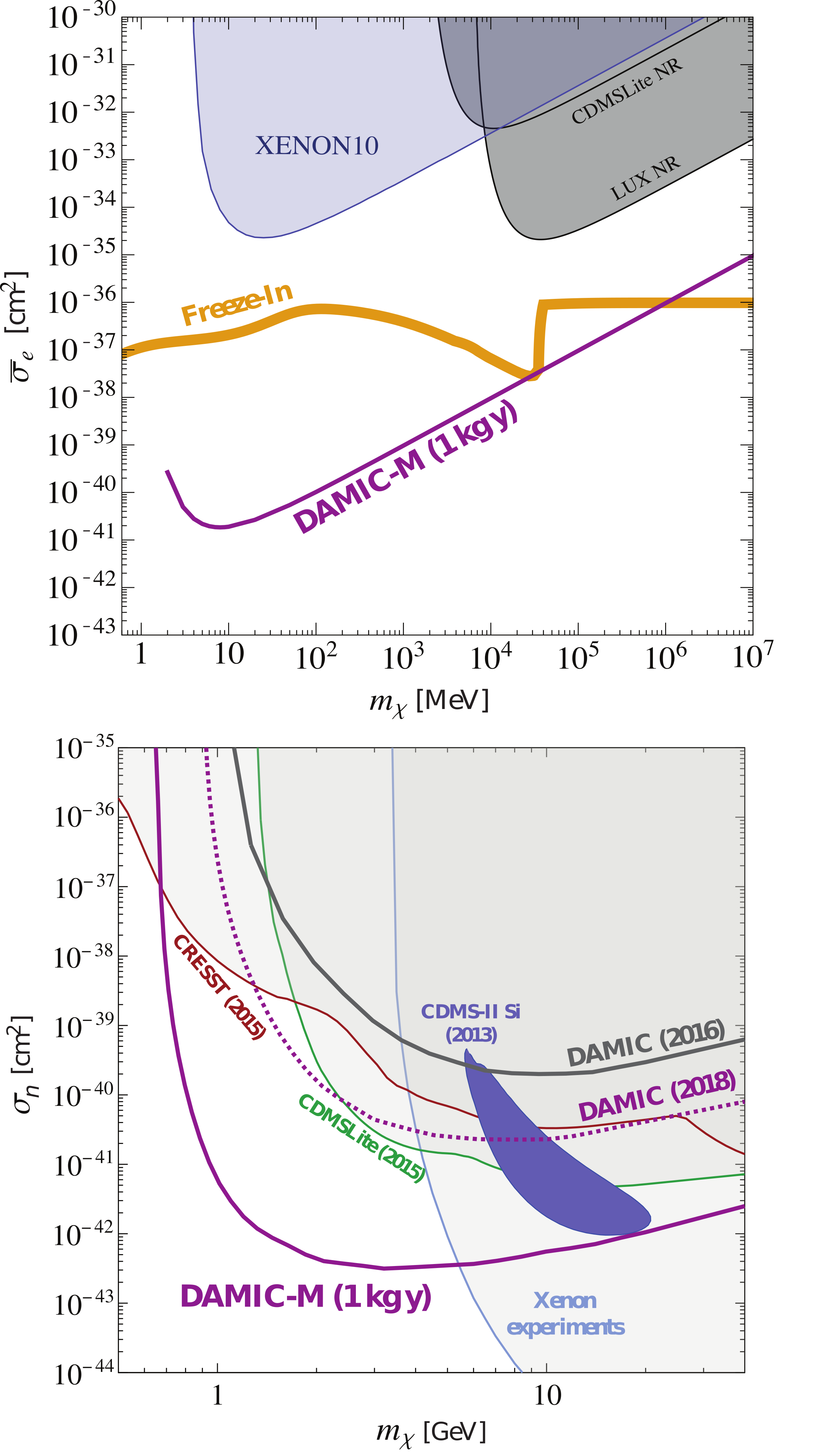}\\
    \caption{Expected sensitivity of DAMIC-M for DM-electron for a light dark photon
        mediator (top panel) and WIMP-nucleon spin-independent scattering 
        (bottom panel).
        Exclusion limits from other dark matter searches are shown for comparison.}
        \label{fig:DMExpSens}
\end{figure}

The sensitivity of DAMIC-M to the hidden sector for an integrated exposure of one kg-year is
    shown in Fig.~\ref{fig:DMExpSens}. 

DAMIC-M will pioneer the low-mass dark-matter searches with unprecedented sensitivity to DM-electron
scattering and hidden-photon DM, by improving by orders of magnitude the sensitivity to
the ionization signal from the scattering of dark matter particles with valence electrons.
Under these conditions, DAMIC-M will be able to progress furhter in the search for low-energy
    dark matter particles, including the GeV-scale WIMPS (left panel on the Figure), the
    hidden-photon, and to probe a large region of parameter space for dark matter particles in the
    \emph{hidden sectors} (not directly coupling with the ordinary matters) and having masses from 1
    MeV/c$^{2}$ to 1 GeV/c$^{2}$ (right panel on the Figure).

\bibliography{vci2019_789} 
\end{document}